\documentclass[aip,amsmath,amssymb]{revtex4-1}
\usepackage[dvipdfmx]{graphicx}
\usepackage{color}	% required for `\textcolor' (yatex added)
\begin{document}
\title{Impact of electrodes on the extraction of shift current from a ferroelectric semiconductor SbSI}

\author{M. Nakamura}
\email{masao.nakamura@riken.jp}
\affiliation{RIKEN Center for Emergent Matter Science (CEMS), Wako, 351-0198, Japan}
\affiliation{PRESTO, Japan Science and Technology Agency (JST), Kawaguchi, 332-0012, Japan}
\author{H. Hatada}
\affiliation{Department of Applied Physics and Quantum Phase Electronics
Center (QPEC), University of Tokyo, Tokyo 113-8656, Japan}
\author{Y. Kaneko}
\affiliation{RIKEN Center for Emergent Matter Science (CEMS), Wako, 351-0198, Japan}
\author{N. Ogawa}
\affiliation{RIKEN Center for Emergent Matter Science (CEMS), Wako, 351-0198, Japan}
\affiliation{PRESTO, Japan Science and Technology Agency (JST), Kawaguchi, 332-0012, Japan}
\author{Y. Tokura}
\affiliation{RIKEN Center for Emergent Matter Science (CEMS), Wako, 351-0198, Japan}
\affiliation{Department of Applied Physics and Quantum Phase Electronics
Center (QPEC), University of Tokyo, Tokyo 113-8656, Japan} 
\author{M. Kawasaki}
\affiliation{RIKEN Center for Emergent Matter Science (CEMS), Wako, 351-0198, Japan}
\affiliation{Department of Applied Physics and Quantum Phase Electronics
Center (QPEC), University of Tokyo, Tokyo 113-8656, Japan} 
\date{\today}

\begin{abstract}
Noncentrosymmetric bulk crystals generate photocurrent without any bias voltage. One of the dominant mechanisms, shift current, comes from a quantum interference of electron wave functions being distinct from classical current caused by electrons' drift or diffusion. The dissipation-less nature of shift current, however, has not been fully verified presumably due to the premature understanding on the role of electrodes. Here we show that the photocurrent dramatically enhances by choosing electrodes with large work function for a $p$-type ferroelectric semiconductor SbSI. An optimized device shows a nearly constant zero-bias photocurrent despite significant variation in photocarrier mobility dependent on temperature, which could be a clear hallmark for the dissipation-less nature of shift current. Distinct from conventional photovoltaic devices, the shift current generator operates as a majority carrier device. The present study provides fundamental design principles for energy-harvesting and photo-detecting devices with novel architectures optimal for the shift current photovoltaic effect.
\end{abstract}

\maketitle
Bulk photovoltaic effect (BPVE), the generation of steady-state photocurrent in noncentrosymmetric bulk crystals in the absence of external voltage application, has been attracting renewed interest~\cite{Butler2015,Paillard2016,Lopez-Varo2016}. The recent development of unprecedented polar semiconductors with large photoresponses to visible and infrared light triggered a movement to reappraise the potentials of BPVE as a new architecture for energy-harvesting and photo-detecting devices~\cite{Yang2009,Grinberg2013,Wu2016,Cook2017}. In particular, BPVE can output a large photovoltage far exceeding the bandgap~\cite{Glass1974}, which is a great advantage for solar cells in terms of gaining high power conversion efficiency using narrow bandgap materials. 

The origin of the BPVE has been a long-standing issue of controversy. It is recognized that the injection current (or sometimes called the ballistic current) and the shift current are the two dominant possible origins~\cite{Sturman1992,Sipe2000}. Both of them are current originating from the second-order nonlinear optical response. The origin of the injection current is the asymmetry in the injection rate of non-thermalized carriers at time-reversed momenta in Brillouin zone~\cite{Sturman1982}. On the other hand, the shift current does not require the broken time-reversal symmetry. It arises from a unidirectional shift of electron positions associated with the asymmetric optical charge transfer. The distance and direction of the charge shift, called the shift vector, is proportional to the difference in the Berry connections between the Bloch wave functions of the ground and excited states, indicating that the shift current is purely quantum-mechanical effect~\cite{Baltz1981,Young2012,Tan2016}. The shift vector is closely related to electric polarization. In the contemporary view, the electric polarization is given by Berry connections of occupied bands, which was originally formulated based on the quantum-mechanical description of the pyroelectric current~\cite{King-Smith1993,Resta1994}. Therefore, in a naive view, the shift current can be regarded as a displacement current continuously pumped by light. The description of the shift current using the Floquet formalism has rendered its topological aspects much clearer~\cite{Morimoto2016}. 

The shift current has a dissipation-less nature in the sense that it is independent of the carrier mobility and robust against the impurity scattering~\cite{Morimoto2018}. This is a stark difference from conventional photocurrent via the electrons' drift and diffusion processes typically observed in photoexcited $p$-$n$ junctions. The injection current is also limited by the mean free path of non-thermalized carriers, although its mobility is much larger than that of thermalized carriers~\cite{Sturman1982,Astafiev1988,Gu2017}. Therefore, the mobility-independent zero-bias photocurrent can be clear evidence for the shift current. However, previous studies have not been able to show this behavior in experiments~\cite{Nakamura2017,Ogawa2017}. For instance, in the case of a ferroelectric organic charge-transfer complex TTF-CA, zero-bias photocurrent takes a maximum near the ferroelectric transition temperature and rapidly decreases at low temperature~\cite{Nakamura2017}. This is inconsistent with the picture of shift current which expects rather temperature-independent zero-bias photocurrent. The reason for this inconsistency is assigned to the formation of the Schottky barrier at the sample/electrode interface~\cite{Nakamura2017}, but there has been no direct evidence. 

In this study, we report the impact of electrode choice upon the extraction of shift current from a representative ferroelectric semiconductor, antimony sulphoiodide (SbSI). The crystal structure of SbSI is shown schematically in Fig.~1(a)~\cite{Momma2011}. It consists of double chains extended one-dimensionally along the c axis. The ferroelectric state appears at a Curie temperature ($T_{\mathrm{C}}$) of 295~K with a large spontaneous polarization ($P_{\mathrm{s}}=20~\mu\mathrm{Ccm}^{-2}$) along the chain direction~\cite{Fatuzzo1962}. The polarization arises from the displacement of I$^{-}$ and S$^{2-}$ anions and Sb$^{3+}$ cations toward opposite directions (Fig.~1(b))~\cite{Kikuchi1967,Amoroso2016}. The theoretically estimated Born effective charges for constituent ions are significantly deviated from the nominal ionic valences~\cite{Amoroso2016}, indicating a sizeable contribution of the covalency to the polarization, which is crucial for the generation of large shift current~\cite{Tan2016, Fregoso2017}. Furthermore, SbSI has strong optical absorption for visible light owing to its narrow optical bandgap of about 2.0~eV~\cite {Novak2013,Grekov1970}. The narrow bandgap, as well as high $T_{\mathrm{C}}$ and large polarization, make SbSI a quite attractive candidate for a solar cell application~\cite{Butler2015,Nitsche1960,Nie2017}. By optimizing the electrodes, we observed photocurrent whose amplitude is independent of the carrier mobility. This is a clear evidence for the dissipation-less nature expected for shift current. We have also revealed that photocurrent is driven by majority carrier, which is in sharp contrast to that in conventional photovoltaic devices, where minority carrier plays an important role.

Needle-like single crystals of SbSI were grown by physical vapor transport technique with typical dimensions of $0.3\times0.3 ~\mathrm{mm}^{2}$ in the cross-section and 5~mm in length. The longitudinal direction of the crystals corresponds to the $c$ axis. Various electrode materials were deposited on the crystals to test the electrode dependence of the photovoltaic effect along the c axis direction as depicted in Fig.~1(c). Here we choose Pt, Au, Ag, Al, and Ca covered with Al (Al/Ca) as metals for the electrode. We also prepared bilayer electrodes consisting of Al with a 1~nm-thick tunneling barrier of LiF (Al/LiF), which is a representative low-work-function electrode employed for an electron-injection layer in organic light-emitting diodes~\cite{Hung1997,Prada2008}. The gap between both electrodes was defined to be 0.45~mm using a stencil mask. Measurements of photocurrent and photovoltage were carried out with the illumination of light from a solar simulator (0.1~Wcm$^{-2}$) which covers the whole sample area between the electrodes as illustrated in Fig.~1(c).

Figure~1(d) shows the temperature dependence of zero-bias photocurrent measured in warming processes after cooling the sample under an electric field of 2.2~kVcm$^{-1}$ to align the polarization in one direction. The magnitude of photocurrent is normalized by the width of the samples. The sharp photocurrent peaks seen at around $T_{\mathrm{C}} = 292~\mathrm{K}$ indicates the output of pyroelectric current upon the ferroelectric transition. Below $T_{\mathrm{C}}$, we observed zero-bias photocurrent whose temperature dependence and magnitude strongly depend on the electrode materials. The samples with large-work-function electrodes like Pt and Au show sizable zero-bias photocurrent which persists down to low temperatures. On the contrary, in the samples with small-work-function electrodes like Al/Ca and Al/LiF, the zero-bias photocurrent is small and rapidly decreases as lowering the temperature. As clearly seen in the work-function dependence of the magnitude of zero-bias photocurrent at 200~K as plotted in Fig.~1(e), the zero-bias photocurrent monotonously increases as increasing the work function of the electrodes.

SbSI is considered to be a $p$-type semiconductor with a large Fermi energy ($E_{\mathrm{F}}$)~\cite{Amoroso2016, Nie2017, Butler2016}. The valence-band maximum (VBM) of SbSI is located 5.5~eV with respect to the vacuum level, and $E_{\mathrm{F}}$ locates 0.5~eV above VBM as shown in the inset of Fig.~1(e). The trend in Fig.~1(e) indicates that the closer the energy level of the electrode to $E_{\mathrm{F}}$, the larger zero-bias photocurrent. This tendency is quite reasonable if the resistance is measured in dark condition because the formation of the Schottky barrier is dependent on the relative position of $E_{\mathrm{F}}$ and work function. The fact that this tendency is valid even for photocurrent indicates that the majority carriers (holes) in SbSI play a dominant role in photovoltaic action as confirmed later.

Having identified the largest photocurrent from the sample with Pt electrodes, we measured current ($I$)-voltage ($V$) characteristics at various temperatures under light irradiation as shown in Fig.~2(a). The slope of the $I$-$V$ curve becomes smaller as lowering temperature, indicating the increase in the resistance of the sample under photoirradiation ($R^{\star}_{\mathrm{bulk}}$). We plot in Fig.~2(c) the temperature dependence of $R^{\star}_{\mathrm{bulk}}$ calculated from the slope of the $I$-$V$ curves. It indicates that $R^{\star}_{\mathrm{bulk}}$ changes by about four orders of magnitude between 40~K and 300~K. 

In general, the conductivity ($\sigma$ ) and photoconductivity ($\sigma_{\mathrm{photo}}$) are given by $\sigma=e(p\mu_{h}+n\mu_{e})$  and $\sigma_{\mathrm{photo}}=e(\Delta p\mu_{h}+\Delta n\mu_{e})$, where $\mu_{h}$  ($\mu_{e}$) denotes the mobility of holes (electrons), $p$ ($n$) the density of free holes (electrons) in the dark condition, and $\Delta p$ ($\Delta n$) the change in the density of free holes (electrons) induced by photoexcitation~\cite{Bube1978}. Here we consider only the transport of holes as a dominant contribution to the photocurrent in SbSI. $\Delta p$ and $\mu_{h}$  is proportional to the life time ($\tau_{\mathrm{life}}$) and the scattering time ($\tau_{\mathrm{scat}}$), respectively. In SbSI, $\tau_{\mathrm{life}}$ is about 10$^{-9}$~s which has been evaluated from time-resolved photoluminescence~\cite{Brandt2017}, whereas $\tau_{\mathrm{scat}}$ is at most $3\times10^{-14}$~s which is estimated using the theoretically predicted effective mass of hole $m_{h}^{\ast}=0.57$~\cite{Butler2016} and assuming $\mu_{h}=100~\mathrm{cm}^{2}V^{-1}s^{-1}$. Since $\tau_{\mathrm{life}}\gg\tau_{\mathrm{scat}}$, the temperature dependence of $R^{\star}_{\mathrm{bulk}}$  can be attributed to that of $\mu_{h}$ . The observed exponential increase in $R^{\star}_{\mathrm{bulk}}$ (decrease in $\sigma_{\mathrm{photo}}$) with lowering temperature indicates that hopping-like carrier transport dominates the conduction in SbSI. Contrary to the large temperature variation in $\mu_{h}$, the zero-bias photocurrent, which is the intersections of $I$-$V$ curves with the vertical axis in Fig.~2(a), remains almost constant value. The observed mobility-independent zero-bias photocurrent is a clear hallmark of the shift current. Neither the injection current nor the drift-diffusion current arising from the internal electric field across the absorber layer can be the origins for it because both of them are affected by scatterings and should critically decrease in lowering temperature.

While the zero-bias photocurrent is nearly temperature independent, the open-circuit photovoltage ($V_{\mathrm{OC}}$), which is the voltage needed to set $I=0$ condition, increases towards low temperature up to 70~V (Fig.~2(b)). The divergent increase of $V_{\mathrm{OC}}$ starts at around 100~K as can be seen in Fig.~2(d). Since the $I$-$V$ curves are almost linear, it is evident that the increase of $V_{\mathrm{OC}}$ originates from the enhancement of $R^{\star}_{\mathrm{bulk}}$. As a matter of course, the temperature dependences of $V_{\mathrm{OC}}$ and $R^{\star}_{\mathrm{bulk}}$ scale well with each other as seen in Fig.~2(c). 

A simple equivalent circuit displayed in the inset of Fig.~2(c) well explains above characteristics~\cite{Nakamura2017}. It consists of a shift current generator ($I_{\mathrm{shift}}$) placed in parallel with $R^{\star}_{\mathrm{bulk}}$ and a contact resistance under photoirradiation ($R^{\star}_{\mathrm{contact}}$) connected in series to them. The electrode dependence displayed in Fig.~1(d) reveals that $R^{\star}_{\mathrm{contact}}$ can be almost eliminated by using the Pt electrodes. In this case, the $R^{\star}_{\mathrm{contact}}$ does not hamper the extraction of shift current. In the open-circuit condition, the shift current is in balance with the drift current driven by $V_{\mathrm{OC}}$. Thus, the output voltage is larger as $R^{\star}_{\mathrm{bulk}}$ increases. Accordingly, materials with lower carrier mobility, such as flat band systems, can show higher power conversion efficiency in shift current photovoltaics~\cite{Morimoto2018}, which is quite contrastive to conventional photovoltaics where the mobility is critical for the conversion efficiency.  We have also measured the light-polarization dependence of zero-bias current~\cite{Sotome2018}. The observed polarization dependence is consistent with the nonlinear conductivity tensor for the point group of SbSI ($C_{2v}$)~\cite{Kikuchi1967}, giving a corroborative evidence for the shift current.

In conventional photovoltaic devices, photocurrent arises from the transport of photogenerated electrons and holes into opposite electrodes. Thus, two electrodes should have different characters: one is a large-work-function electrode for the extraction of holes (anode) and the other a small-work-function electrode for electrons (cathode). We examined the effect of such asymmetric electrode structure on shift current. As given in the inset schematic of Fig.~3(c), we employ Pt (Al/LiF) for the anode (cathode), which we call Pt//Al/LiF sample hereafter. We also examined the photovoltaic properties of Pt//Pt and Al/LiF//Al/LiF samples with symmetric electrode structures. 

For symmetric Pt//Pt and Al/LiF//Al/LiF samples, the zero-bias photocurrent is almost symmetrically reversed by reversing the polarization direction as shown in Figs.~3(a) and 3(b). On the other hand, the Pt//Al/LiF sample exhibits a positive sign irrespective of the polarization direction at a high temperature, although sign reversal occurs at a low temperature as shown in Fig.~3(c). The band bending in a semiconductor in contact with a metal electrode can generate classical photocurrent via the drift-diffusion process. This current should vanish in the symmetric electrode samples due to the opposite band bending at the two electrodes (Figs.~3(e)-1 and 3(e)-2), while it can be finite in the asymmetric electrode samples (Fig.~3(e)-3). The polarization-independent photocurrent observed in the high-temperature region in the Pt//Al/LiF sample is ascribed to the drift-diffusion current. The observed positive sign of the photocurrent is consistent with that expected from the profile of the band bending. The zero-bias photocurrent for the Pt//Al/LiF sample can be deconvoluted into the classical drift-diffusion current and the quantum-mechanical shift current by symmetrizing and antisymmetrizing the current for opposite polarization states, respectively, as shown in Fig.~3(d). The drift-diffusion current rapidly disappears as lowering the temperature, which is definitely caused by the reduction of photocarrier mobility. In contrast, the shift current remains constant or even increases toward the lowest temperature. The increase of the shift current below 150~K probably stems from the divergent increase in $R^{\star}_{\mathrm{bulk}}$, because the output photocurrent for the equivalent circuit in the inset of Fig.~2(c) is given by $R^{\star}_{\mathrm{bulk}}/(R^{\star}_{\mathrm{bulk}}+R^{\star}_{\mathrm{contact}})\times I_{\mathrm{shift}}$, which approaches to the value of $I_{\mathrm{shift}}$ when $R^{\star}_{\mathrm{bulk}}\gg R^{\star}_{\mathrm{contact}}$.

The magnitude of the zero-bias photocurrent in the Pt//Al/LiF sample (Fig.~3(c)) turns out to be smaller than that of the Pt//Pt symmetric electrode sample (Fig.~3(a)), rather close to that of Al/LiF//Al/LiF symmetric electrode sample (Fig.~3(b)). The result indicates that the asymmetric electrode structure indispensable to extract classical drift-diffusion current is not beneficial but rather harmful for extracting shift current. As can be seen in the band profiles, there is no potential barrier for holes at the interface between SbSi and Pt (Figs.~3(e)-1, 3(e)-3), and holes can smoothly transfer to the electrodes. On the other hand, the potential barrier formed at the interface between SbSI and Al/LiF impedes the transfer of holes and eventually gives rise to the considerable contact resistance (Figs.~3(e)-2 and 3(e)-3). Thus, the amplitude of zero-bias photocurrent is in the order of Pt//Pt $>$ Pt//Al/LiF $>$ Al/LiF//Al/LiF. From these results, we conclude that majority carriers dominantly contribute to the shift current generation as schematically shown in Fig.~3(e). The majority-carrier operation in the shift current generation is in stark contrast to $p-n$-junction-based photovoltaics those operate as minority-carrier devices.

Besides the robustness against the scatterings, the shift current characteristic features of the ultra-fast response to pulsed light and the low current noise~\cite{Morimoto2018,Sotome2018,Laman2005}, which are substantial advantages not only in energy harvesting devices but also in photodetectors. Toward the development of these new optoelectronic devices based on the shift current mechanism, the requisites for electrode material as well as the unique operation mechanism elucidated in this study will give a guideline to design an optimum device structure.

In conclusion, we have demonstrated that the output of shift current is strongly dependent on the work function of the electrode materials. For a $p$-type ferroelectric semiconductor SbSI, large-work-function electrodes are superior to efficiently extract shift current. By using optimum electrodes, we could verify the dissipation-less nature of shift current in the sense that it is not affected by the scattering of carriers, hence that the shift current is driven via quantum-mechanical motion of majority carriers. The present study will promote successive researches to improve the performance of photovoltaic devices as well as explore novel quantum-mechanical phenomena in photovoltaic effect with its origin in the Berry phase of electron wave functions.

\begin{acknowledgments}
We would like to thank N. Nagaosa, T. Morimoto and M. Sotome for fruitful discussions. This work was partly supported by PRESTO JST (JPMJPR16R5).
\end{acknowledgments}

\newpage
\begin{figure}
\begin{center}
\includegraphics[width=14cm]{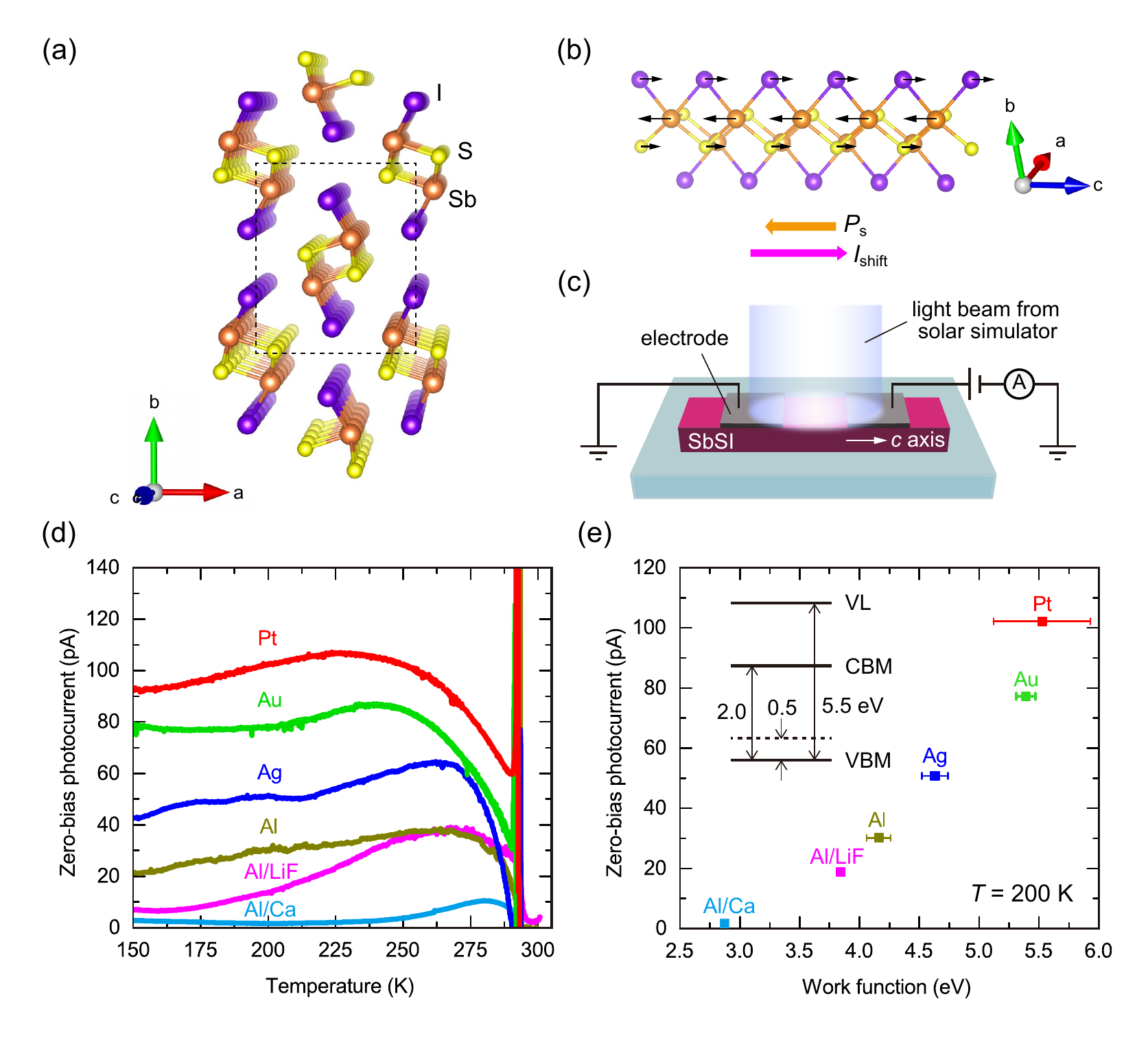}
\end{center}
\caption{(a) Crystal structure of SbSI. (b) Relationship among atomic displacement, spontaneous polarization ($P_{\mathrm{s}}$), and shift current ($I_{\mathrm{shift}}$). (c) A schematic of measurement configuration. Light from a solar simulator is on the entire area between electrodes ($\sim$0.45~mm). (d) Temperature dependence of zero-bias photocurrent for SbSI samples with different electrode materials. Current is normalized by the sample width to a typical value of 0.3~mm. (e) Relation between the magnitude of zero-bias photocurrent at 200~K and work function of electrode. The values of work functions are quoted from Refs.~\onlinecite{Prada2008} and \onlinecite{Lide2012}. Inset shows the band diagram of SbSI. VL, CBM, and VBM denote vacuum level, conduction band minimum, and valence band maximum, respectively.}
\end{figure}

\begin{figure}
\begin{center}
\includegraphics[width=14cm]{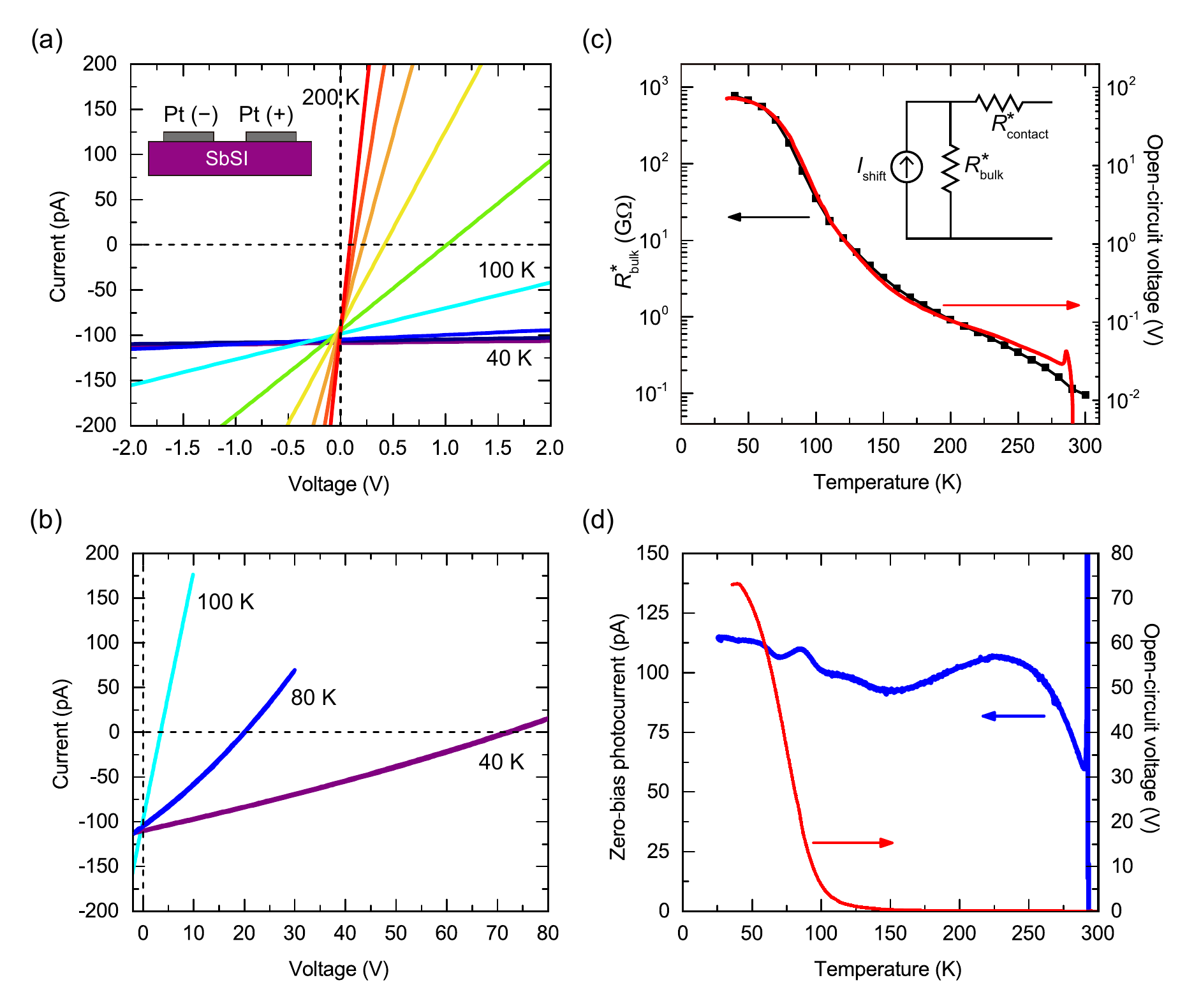}
\end{center}
\caption{(a) Current-Voltage ($I$-$V$) characteristics measured at various temperatures. The intercept points of the $I$-$V$ curves with the vertical and horizontal axes indicate zero-bias photocurrent and open-circuit photovoltage ($V_{\mathrm{OC}}$), respectively.  The inset shows a schematic of measured sample. (b) $I$-$V$ characteristics with wider voltage range in low temperature region. (c) Temperature dependence of bulk resistance under photoirradiation ($R^{\star}_{\mathrm{bulk}}$) and $V_{\mathrm{OC}}$. $R^{\star}_{\mathrm{bulk}}$ is derived from the slope of the $I$-$V$ curves shown in (a). The inset is an equivalent circuit for a shift current generator. $R^{\star}_{\mathrm{contact}}$ denotes the contact resistance, for example, caused by Schottkey barrier formation at sample/electrode interfaces. Since $R^{\star}_{\mathrm{contact}}$ is negligibly small in the Pt//Pt sample, the temperature dependence of $V_{\mathrm{OC}}$ scales with that of $R^{\star}_{\mathrm{bulk}}$. (d) Temperature dependence of zero-bias photocurrent and $V_{\mathrm{OC}}$ measured during increasing the temperature.}
\end{figure}

\begin{figure}
\begin{center}
\includegraphics[width=16cm]{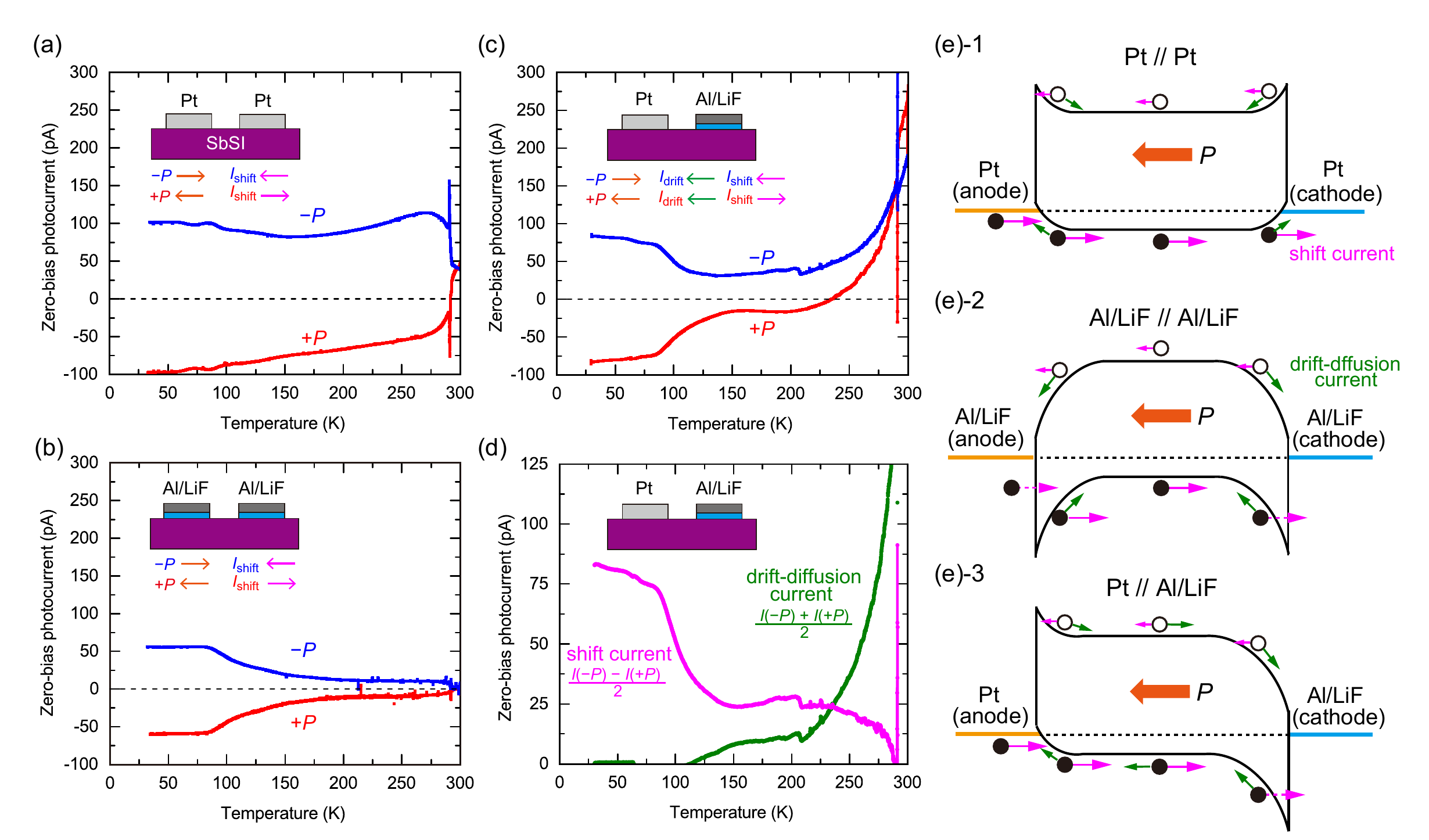}
\end{center}
\caption{(a), (b), (c) Temperature dependence of zero-bias photocurrent for SbSI with Pt//Pt symmetric electrodes (a), Al/LiF//Al/LiF symmetric electrodes (b) and Pt//Al/LiF asymmetric electrodes (c). Photocurrent was measured after the sample was poled in positive ($+P$) and negative ($-P$) directions. The insets are schematics of the relation between polarization and photocurrent. (d) Temperature dependence of drift-diffusion current and shift current in the Pt//Al/LiF sample. Each component is derived by symmetrizing and asymmetrizing the zero-bias photocurrent data for opposite polarizations. (e) Possible band profiles and mechanisms of photocurrent generation in the three electrode structures in the positively poled condition. The pink arrows denote the displacement of photogenerated electrons and holes by the shift-current mechanism. The difference in the length of the arrows between electrons and holes indicates the difference in their contributions to shift current generation. The dashed arrows indicate that the transport of holes via the shift-current mechanism is impeded by the Schottky barrier at the interfaces between Al/LiF and SbSI. The green arrows denote the motion of photocarriers due to the drift-diffusion mechanism.}
\end{figure}
\end{document}